\documentstyle[preprint,prl,aps]{revtex}
\begin{document}
\draft
\title{Kinetic Theory Approach to the SK Spin Glass Model with 
Glauber Dynamics}
\author{Grzegorz Szamel}
\address{Department of Chemistry,
Colorado State University, Fort Collins, CO 80523, USA
}
\date{\today}
\maketitle

\begin{abstract}
I present a new method to analyze Glauber dynamics of
the Sherrington-Kirkpatrick (SK) spin glass model. The method is
based on ideas used in the classical kinetic theory of fluids.
I apply it to study spin correlations in the high temperature
phase ($T\ge T_c$) of the SK model at zero external field. The zeroth 
order theory is equivalent to a disorder dependent local equilibrium 
approximation. Its predictions agree well with
computer simulation results. The first order theory 
involves coupled evolution equations for the spin correlations
and the dynamic (excess) parts of the local field distributions.
It accounts qualitatively for the error made in the zeroth 
approximation. 
\end{abstract}


\narrowtext
The dynamical properties of the Sherrington--Kirkpatrick (SK) 
spin glass model \cite{SK}
have been a subject of continuous interest in recent years \cite{rev}. 
However, almost all the theoretical studies considered
Langevin dynamics of the soft-spin version of the SK model
\cite{SZ,S,CK}.
The soft-spin version, while showing very interesting 
dynamical properties \cite{CK}, lacks the original 
motivation of the SK model: neither its statics nor its dynamics 
is exactly solvable \cite{comment0}. 
The Glauber dynamics of the 
SK model was studied by Sommers \cite{Sommers}.
He recovered the results found previously for the Langevin dynamics.
Sommers' method was criticized by \L usakowski \cite{Lus}
and its validity is uncertain \cite{comment}.
Recently a novel approach to Glauber dynamics of 
spin glasses \cite{CS,CS2} has been proposed 
by Coolen, Sherrington, and coworkers (CS).
The simple version of their theory \cite{CS} describes 
very well the order parameter 
{\em flow direction} above the de Almeida-Thouless (AT) \cite{AT}
line but misses the {\em slowing down} 
which sets in when the former line is approached from above. 
The more advanced version 
\cite{CS2} agrees well with the simulation data for 
short times but it remains to be seen 
whether it predicts divergent relaxation times at and below the AT line.

Here I reconsider the Glauber dynamics of the SK model. The original 
motivation for this work was to improve the simple CS theory 
\cite{CS}. 
However, the resulting method is very different from that of CS.  

CS tried to derive a general description of the SK spin glass
dynamics. The theory presented here is 
more restricted: I study time-dependent spin correlations 
{\em in equilibrium} in the high temperature phase 
($T\ge T_c$) at zero external field. 
The main motivation is simplicity: 
it is possible to derive explicit results for these 
correlations, and it is easy to perform accurate
computer simulations that allow testing the theoretical predictions. 

Following an approach used in kinetic theory \cite{LPS},
I express the correlation functions in terms of 
a distribution that
satisfies the master equation and a specific initial condition. 
Next I propose a series of approximations for this distribution 
that are motivated by the approximations used in the kinetic theory 
\cite{Resetal}. 
Successive approximations gradually include {\em dynamic}
many-spin correlations. The static correlations are retained
at every step.

The approximations are formulated for a given sample of
the coupling constants. The averaging over the samples 
is postponed until after the resulting evolution equations are solved.

The simplest (0th order) approximation 
is equivalent to a {\em disorder dependent} version of the local 
equilibrium approximation \cite{Kawasaki}.
It leads to very simple equations of motion for the
spin correlations: 
the relaxation matrix is a product of a relaxation rate 
(kinetic coefficient), $\tau^{-1}$, 
that is finite at the transition temperature, $T_c$, 
and an inverse matrix of equilibrium spin correlations, $A_{ij}$,
or the Hessian of the Thouless-Anderson-Palmer (TAP)
\cite{TAP} free energy. The Hessian
acquires zero eigenvalues at $T_c$ \cite{BM}.
This results in a mean-field-like critical slowing down of the 
time-dependent correlations when $T_c$ is approached from above 
and an algebraic decay $\sim t^{-1/2}$ at $T_c$. A comparison with 
the simulation data shows that the the zeroth order 
approximation is surprisingly accurate. 

The first order approximation takes into account {\em dynamic}
correlations between spins and the distributions of the local fields
acting on these spins: it includes time-delayed Onsager
reaction fields. 

The first order approximation accounts qualitatively for the error
made in the zeroth order: the predicted {\em difference} between
the full correlations and the zeroth order approximation 
is about 40\% of the simulation result.

I now sketch the derivation of the results. I consider the Glauber 
dynamics for the SK model of a spin glass. The time evolution is given by
the master equation for the  spin probability distribution
$P( {\mbox{\boldmath$\sigma$}};t)$,
\begin{equation}\label{Smol}
\partial P({\mbox{\boldmath$\sigma$}};t)/\partial t = - 
\sum_i (1-S_i) w_i({\mbox{\boldmath$\sigma$}}) 
P({\mbox{\boldmath$\sigma$}};t).
\end{equation}
Here ${\mbox{\boldmath$\sigma$}}\equiv \{\sigma_1, ..., \sigma_N\}$ 
denotes the spin
configuration, $S_i$ is the spin-flip operator, $S_i \sigma_i =
-\sigma_i$, and $w_i({\mbox{\boldmath$\sigma$}})$
is the transition rate, $w_i({\mbox{\boldmath$\sigma$}})=
(1-\sigma_i \tanh(\beta h_i))/2$, with
$h_i$ being a local magnetic field acting on the $i$th spin,
$h_i = \sum_{j\neq i} J_{ij} \sigma_j$.  The $J_{ij}$ are 
the exchange coupling constants that are quenched random variables
distributed according to the symmetric distribution
$P(J_{ij})\sim \exp(-J_{ij}^2/(2J^2/N))$.

I study the  time-dependent correlations of the total 
magnetization in equilibrium, 
$(1/N)\left[\left< m(t) m(0)\right>_{eq}\right]$. Here 
$m(t)=\sum_i \sigma_i(t)$ 
is the fluctuation of the magnetization (for $T\ge T_c$ at zero external
field $\left<\sigma_i\right>_{eq}\equiv 0$),
the angular brackets $< ... >_{eq}$ denote
the equilibrium ensemble average, and
the square brackets $[...]$ denote the sample averaging
over the distribution of $J_{ij}$'s.

I perform the sample averaging at the very last stage of the analysis.
Therefore for the most part I deal with sample dependent
quantities like $\left<\sigma_i(t) m(0)\right>_{eq}$.
This is analogous to the TAP \cite{TAP} analysis of the equilibrium SK 
model and to early work \cite{SK,KF} on the Glauber dynamics of the SK 
model. It is different from the CS approach and also from most of the 
other approaches to both Langevin \cite{SZ,S,CK} and 
Glauber \cite{Sommers} dynamics. 

The correlations $\left<\sigma_i(t) m(0)\right>_{eq}$
are defined in terms of a conditional distribution 
$P({\mbox{\boldmath$\sigma$}};t|{\mbox{\boldmath$\sigma'$}})$ 
and the equilibrium distribution
$P_{eq}({\mbox{\boldmath$\sigma$}})$ \cite{Glauber}.
I define a distribution $P_m({\mbox{\boldmath$\sigma$}};t)$,
\begin{equation}\label{Pmdef}
P_m({\mbox{\boldmath$\sigma$}};t) \equiv 
\sum_{{\mbox{\boldmath$\sigma'$}}} 
P({\mbox{\boldmath$\sigma$}};t|{\mbox{\boldmath$\sigma'$}}) 
(\sum_j \sigma_j') P_{eq}({\mbox{\boldmath$\sigma'$}}).
\end{equation}
The distribution $P_m$ satisfies the master equation (\ref{Smol})
and the initial condition \cite{comment2}
$$P_m({\mbox{\boldmath$\sigma$}};t\!=\!0) =  
P_{eq}({\mbox{\boldmath$\sigma$}})\sum_i \sigma_i.$$ 

The time-dependent spin correlations in equilibrium 
$\left<\sigma_i(t) m(0)\right>_{eq}$ can be calculated
as averages over $P_m$,
\begin{equation}\label{corr}
\left<\sigma_i(t) m(0)\right>_{eq} = 
\sum_{{\mbox{\boldmath$\sigma$}}} \sigma_i 
P_m({\mbox{\boldmath$\sigma$}};t) =
\left<\sigma_i \right>(t).
\end{equation}
Hereafter $\left< ... \right>(t)$ denotes average over 
the time-dependent distribution $P_m$. In the following I propose
a series of approximations for this distribution.

In the zeroth approximation I
assume that $P_m(t)$ can be expressed in terms of the single-spin
averages, $\left< \sigma_i \right>(t)$. More precisely, I assume that 
$P_m$ has the same form as an equilibrium distribution for the
system in an external field with the field chosen in such a way
that the single spin averages have correct values. Explicitly,
\begin{equation}\label{Pmt}
P_m({\mbox{\boldmath$\sigma$}};t) \approx 
P_{eq}({\mbox{\boldmath$\sigma$}}) \sum_i \sigma_i b_i(t),
\end{equation}
where fields $b_i(t), i=1, ...$ satisfy the following equations,
\begin{equation}\label{cond}
\left<\sigma_i \right>(t) = 
\sum_k \left<\sigma_i   \sigma_k \right>_{eq} b_k(t).
\end{equation}
Solving Eq. (\ref{cond}) for the $b_k(t)$ I get 
\begin{equation}\label{Pmt2}
P_m({\mbox{\boldmath$\sigma$}};t) \approx 
P_{eq}({\mbox{\boldmath$\sigma$}}) \sum_{ij} \sigma_i A_{ij} 
\left<\sigma_j \right>(t).
\end{equation}
Here the matrix $A_{ij}$ is the inverse of the matrix of
the equilibrium spin correlations, $\sum_j A_{ij} 
\left<\delta\sigma_j \delta\sigma_k\right>_{eq} =\delta_{ik}$. 
Note that $A_{ij}$ is identical to the Hessian of the TAP free energy 
\cite{TAP}. 

The ansatz (\ref{Pmt2}) is similar to the local equilibrium approximation  
introduced by Kawasaki \cite{Kawasaki}. The new element of this work is to 
use the local equilibrium approximation for the 
{\em disorder dependent} distribution $P_m$.

To derive the equations of motion for the spin averages I start from 
the exact evolution equations,
\begin{equation}\label{hierarchy1}
\partial \left<\sigma_i \right>(t) /\partial t =
-\left<\sigma_i \right>(t) +  \left<\tanh(\beta h_i)\right>(t).
\end{equation}
Then I use ansatz (\ref{Pmt2}) to calculate the averages at 
the right-hand-side (RHS) of Eqs. (\ref{hierarchy1}) and obtain 
\begin{equation}\label{eom1}
\frac{\partial \left<\sigma_i \right>(t)}{\partial t} =
\left( -1 + \left<\tanh(\beta h_i) \sigma_i \right>_{eq} \right)
\sum_j A_{ij} \left<\sigma_j \right>(t).
\end{equation}

According to Eqs. (\ref{eom1}) the dynamics of the spin correlations 
follows a van Hove mean-field-like picture: the relaxation matrix is
a product of the relaxation rate,
$\tau^{-1} =1-\left< \tanh(\beta h_i) \sigma_i \right>_{eq}$,
and the inverse matrix of the spin correlations (Hessian), $A_{ij}$.
Each of Eqs. (\ref{eom1}) contains
an Onsager correction term \cite{SK} 
that has been introduced phenomenologically 
in early works \cite{SK,KF}. Within the zeroth order theory the 
correction term is {\em instantaneous}: 
the reaction field at a given time depends on the value 
of the spin average at the same time.

The relaxation rate can be calculated with the help of the equilibrium
probability distribution of the local fields, $P_{eq}(h)$ 
\cite{local_fields}. Numerical evaluation 
shows that at $T_c$ the relaxation rate is finite.
On the other hand the Hessian,  
$A_{ij}$, acquires zero eigenvalues at the transition temperature 
\cite{BM} and this fact leads to a mean-field-like critical slowing down
as $T_c$ is approached from above. Moreover, at $T_c$ I obtain 
asymptotically $ \left[\left<\sigma_i \right>(t)\right] \sim t^{-1/2}$.

In the high temperature phase at zero external field
the Hessian is known explicitly:
\begin{equation}\label{Hessian}
A_{ij} = 
- \beta J_{ij} + \delta_{ij} (1+ (\beta J)^2).
\end{equation}
It follows that the evolution equations 
(\ref{eom1}) are almost identical to those derived in the original SK
paper \cite{SK}.
The solution has the same form as the solution of the SK 
equations if the time scale of SK is rescaled by factor $\tau$.

In Fig. 1 I compare predictions of the zeroth order theory with numerical
simulations of the SK model at the transition temperature. 
10 samples of $N=10000$ spins each have been simulated using 
algorithm of Mackenzie and Young \cite{McKY}. Very long equilibration 
time of 10000 Monte Carlo steps per spin (MCS) was used. Subsequently
the data for the time-dependent correlation function 
$(1/N) \sum_i \left< \sigma_i(t) \sigma_i(0) \right>_{eq}$ 
were collected \cite{comment4}
and averaged over different time origins \cite{AllenT}.
The figure indicates that the zeroth order theory is quite accurate:
its predictions differ from the simulation data by less than 11\%.
In Fig. 1 I also plot predictions of the second order Sommers theory.
They were obtained by solving explicitly Eq. (18) of Ref. \cite{Sommers},
using the fluctuation-dissipation theorem to get
the Laplace transform of the correlation function, 
and finally inverting the Laplace transform numerically \cite{Laplace}.

In Fig. 2 I plot the difference between the simulation data and 
the predictions of the zeroth order theory. It is clear that the zeroth 
order approximation is not exact. This fact can also be seen 
from an analysis of the short time behavior of the spin
correlations: the zeroth order theory reproduces exactly
the first time derivative at $t\!=\!0$ but {\em not} the second and 
higher order derivatives.

To improve upon the zeroth order theory it is necessary to go beyond
the local equilibrium approximation and include {\em dynamic}
correlations \cite{Resetal,SL}. It follows from the physics of the 
SK model and from the analysis of the short time expansion of the 
time-dependent spin correlations that the first additional set of 
variables to be included are the dynamic (excess) parts of the 
local field distributions, $\delta P_i(h;t)$ \cite{SCremark}. 
They are defined as the differences between the true distributions
and their values in the local equilibrium ensemble (\ref{Pmt2}),
\begin{eqnarray}\label{deltaP}
\delta P_i(h;t) &=& \left< \delta(h-h_i)  \right>(t) 
\nonumber \\
& - & \sum_{jk} \left< \delta(h-h_i) \sigma_j\right>_{eq} A_{jk} 
\left<\sigma_k\right>(t).
\end{eqnarray}
At $t\!=\!0$ the excess parts vanish, $\delta P_i(h;t\!=\!0) = 0$.

To derive equations of motion for the spin averages and the
excess parts of the local field distributions I need an approximate
expression for the distribution $P_m$ in terms of 
$\left< \sigma_i \right>(t)$ and $\delta P_i(h;t)$. 
I assume that $P_m$ has the following form:
\begin{equation}\label{Pmt3}
P_m({\mbox{\boldmath$\sigma$}};t) \approx 
P_{eq}({\mbox{\boldmath$\sigma$}})
\left( \sum_{ij} \sigma_i A_{ij} \left<\sigma_j\right>(t) 
+ \sum_{ij} \int dh\, \delta^{e}(h-h_i) \int dq\,
C_{ij}(h,q) \delta P_j(q;t) \right).
\end{equation}
Here $\delta^{e}(h-h_i)$ is the microscopic expression for the excess
part of the local field distribution, 
\begin{equation}\label{delta2}
\delta^{e}(h-h_i) = \delta(h-h_i) - 
\sum_{jk} \left< \delta(h-h_i) \sigma_j \right>_{eq} A_{jk} \sigma_k,
\end{equation}
and $C_{ij}(h,q)$ is the inverse ``matrix'' of the correlations
of the excess local field distributions, $\sum_j \int dq\,C_{ij}(h,q)
\left< \delta^{e}(q-h_j) \delta^{e}(p-h_k) \right>_{eq} = 
\delta_{ik} \delta(h-p)$.

The form of the distribution (\ref{Pmt3}) is motivated by 
approximations used in the kinetic theory \cite{Resetal,SL}. 
Briefly, to get (\ref{Pmt3}) I assume that $P_m(t)$ has the same form as 
an equilibrium distribution for the system in the presence of
external perturbations that are chosen in such a way that, at a given time,
the single spin averages and the excess parts of the local field 
distributions are $\left<\sigma_i\right>(t)$ and $\delta P_i(h;t)$, 
respectively.
Now I will show that with the help of (\ref{Pmt3}) one can describe
qualitatively the difference between the predictions of the local
equilibrium approximation and the simulation data. 

First I derive equations of motion for the spin averages. I start from
the exact equations (\ref{hierarchy1}), use (\ref{Pmt3}) to
calculate averages, and obtain the following equations of motion,
\begin{equation}\label{eom2}
\frac{\partial\left<\sigma_i \right>(t)}{\partial t}=
-\frac{1}{\tau}
\sum_j A_{ij} \left<\sigma_j \right>(t) + 
\int dh\,\tanh(\beta h) \delta P_i(h;t).
\end{equation}

Next I derive equations of motion for the excess parts of the
local field distributions. To this end I start from exact evolution
equations,
\begin{equation}\label{hierarchy2}
\frac{\partial\delta P_i(h;t)}{\partial t} = 
\left< \left[ \sum_i  w_i({\mbox{\boldmath$\sigma$}}) (1-S_i) 
\delta^{e}(h-h_i) \right] \right>(t),
\end{equation}
use the distribution (\ref{Pmt3}), and get 
\begin{eqnarray}\label{eom3}
\frac{\partial\delta P_i(h;t)}{\partial t}
&=& 
- \left[ (\beta J)^2 \tanh(\beta h) - \beta h\right] P_{eq}(h) 
\frac{\partial}{\partial t} \left<\sigma_i \right>(t)
\nonumber \\
&&+
\frac{(\beta J)^2}{\tau} \frac{\partial}{\partial \beta h} 
\int dq \,
\left[ P_{eq}(h) \delta(h-q) - P_{eq}(h) P_{eq}(q)\right]
\frac{\partial}{\partial \beta q}
\int dq' \, C_{ii}(q,q') \delta P_i(q';t),
\nonumber \\
\end{eqnarray}
where $P_{eq}(h)$ is the equilibrium local field distribution.
To derive (\ref{eom3}) I keep two-point equilibrium correlations,
{\it e.g.}, $\left< \sigma_i \sigma_j \right>_{eq}$,
but I neglect {\em higher order connected} correlations
involving {\em different} lattice sites, 
{\it e.g.}, $\left< \delta^{e}(h-h_i) \delta^{e}(q-h_j) \right>_{eq}$
for $i\neq j$ \cite{comment5}. 

According to Eqs. (\ref{eom2}) the reaction field consists of two parts:
an instantaneous reaction, proportional to the
spin average at the same time, and a time-delayed reaction.
It follows from (\ref{eom3}) that the delayed reaction field acting on
$i$th spin at a given time depends on the values of 
this spin at earlier times. 

Solving formally Eqs. (\ref{eom3}) and substituting
the result into Eqs. (\ref{eom2}) one gets a set of effective
equations of motion that involve only the single-spin averages. 
These equations have
the same {\em structure} as the memory function equation derived
recently by Kawasaki for dissipative stochastic systems \cite{Kawasaki2}:
the inverse frequency, 
$\tau$, gets renormalized by inclusion of the dynamic correlations.

To derive explicit results for the time-dependent correlations
I solve the integro-differential equation (\ref{eom3}) approximating
$\delta P_i(h;t)$ by a finite sum of basic functions. 

In Fig. 2 I compare theoretical predictions for the difference between 
the full spin correlations and the local equilibrium result 
against the simulation data. 
The agreement is quantitative for short times 
(the first order theory reproduces exactly first two 
time derivatives of the spin correlations at $t\!=\!0$) 
and qualitative at long times. 

It is evident from Fig. 2, and it can be shown theoretically, that
the first order theory is not exact. 
Glauber dynamics of the SK model is more 
complicated than statics: in addition to (possibly time-delayed) Onsager 
reaction fields other {\em dynamic} correlations have to be included.

In summary, I have shown that the ideas of the classical kinetic theory
of fluids can be used to analyze Glauber dynamics of the SK spin glass 
model. The very simple disorder dependent local equilibrium approximation 
leads to quantitatively accurate predictions for the spin correlations, 
at least in the high temperature phase of the SK model.
This fact suggests that {\em a} disorder dependent approximation
might be a good starting
point in the search for a general theory of the SK model dynamics.
Secondly, the error made in the zeroth approximation can be accounted 
for by including time-delayed Onsager reaction fields. The resulting
theory can be improved further by incorporating more complicated
dynamic correlations. 
Finally, the method presented here can be generalized to 
dynamics of neural networks and other Ising-like systems. 

I would like to thank Marshall Fixman for stimulating discussions
and critical comments on the manuscript. This work was partially supported
by NSF Grant No. CHE-9624596.

\begin{figure}
\caption{Spin correlation function at the transition temperature $T_c$.
Crosses: Glauber dynamics simulation data; dashed line: the
zeroth order theory (local equilibrium approximation); 
solid line: the first order approximation; 
dotted line: the second order Sommers' theory.}
\vskip 2ex
\caption{The difference between the full spin correlation function
and the local equilibrium approximation at the transition temperature 
$T_c$. Crosses: Glauber dynamics simulation data; solid line: the 
first order approximation.}
\end{figure}

\begin{references}
\bibitem{SK} S. Kirkpatrick and D. Sherrington, Phys. Rev B {\bf 17},
4384 (1978).
\bibitem{rev} For a review see, e.g., K. Binder and A. P. Young,
Rev. Mod. Phys. {\bf 58}, 801 (1986); 
K H. Fisher and J. A. Hertz, {\it Spin glasses} (Cambridge Univ. Press,
Cambridge, 1991).
\bibitem{SZ} H. Sompolinsky and A. Zippelius, Phys. Rev. Lett. {\bf 47}, 
359 (1981), Phys. Rev. B {\bf 25} (1982).
\bibitem{S} H. Sompolinsky, Phys. Rev. Lett. {\bf 47}, 935 (1981).
\bibitem{CK} L. F. Cugliandolo and J. Kurchan, J. Phys. A {\bf 27}, 
5749 (1994). 
\bibitem{comment0} The results are obtained perturbatively
with respect to the four spin coupling $u$. To recover
the Ising limit one has to let $u$ approach infinity.
In practice, this procedure allows one to analyze the long-time
asymptotic behavior of the spin correlations. It is not well suited
to study the time dependence for all times (even for $T\ge T_c$) 
or to calculate so-called absolute frequency scale.
\bibitem{Sommers} H. J. Sommers, Phys. Rev. Lett. {\bf 58}, 1268 (1987).
\bibitem{Lus} A. \L usakowski, Phys. Rev. Lett. {\bf 66}, 2543 (1991).
\bibitem{comment} Note that in a so-called spherical SK model
the Langevin and Monte Carlo dynamics
lead to {\em different} results [L. L. Bonilla {\it et al.}, 
Europhys. Lett. {\bf 34}, 159 (1996)]. 
\bibitem{CS} A. C. C. Coolen and D. Sherrington, Phys. Rev. Lett. 
{\bf 71}, 3886 (1993), J. Phys. A {\bf 27}, 7687 (1994).
\bibitem{CS2} S. N. Laughton, A. C. C. Coolen, and D. Sherrington, 
J. Phys. A {\bf 29}, 763 (1996).
\bibitem{AT} J. R. L. de Almeida and D. J. Thouless, J. Phys. A {\bf 11},
983 (1978).
\bibitem{LPS} J. L. Lebowitz, J. K. Percus and J. Sykes, 
Phys. Rev. {\bf 188}, 487 (1969). 
\bibitem{Resetal} P. Resibois and J. L. Lebowitz, J. Stat. Phys. {\bf 12}, 
483 (1975); J. B\l awzdziewicz and B. Cichocki, Physica 127A, 38 (1984).
\bibitem{Kawasaki} K. Kawasaki, Phys. Rev. {\bf 145}, 224 (1966); 
{\it ibid}, {\bf 148}, 375 (1966).
\bibitem{TAP} D. J. Thouless, P. W. Anderson and R. G. Palmer,
Phil. Mag. {\bf 35}, 593 (1977).
\bibitem{BM} A. J. Bray and M. A. Moore, J. Phys. C {\bf 12}, 
L441 (1979).
\bibitem{KF} W. Kinzel and K. H. Fisher, Solid State Commun. {\bf 23}, 
687 (1977).
\bibitem{Glauber} R. J. Glauber, J. Math. Phys. {\bf 4}, 294 (1963), 
Eq. (70).
\bibitem{comment2} Note that $P_m(t)$ is normalized to 0 and
therefore, strictly speaking, is not a {\em probability} distribution. 
\bibitem{local_fields} M. Thomsen {\it et al.},
Phys. Rev. B {\bf 33}, 1931 (1986).
\bibitem{McKY} N. D. Mackenzie and A. P. Young, J. Phys. C {\bf 16}, 
5321 (1983).
\bibitem{comment4} To improve the statistics 
I use the identity 
$(1/N)\left[ \left< m(t) m(0) \right>_{eq}\right] = 
\left[\left< \sigma_i(t) \sigma_i(0) \right>_{eq}\right]$.
\bibitem{AllenT} M. P. Allen and D. J. Tildesley, {\it Computer
Simulation of Liquids} (Clarendon Press, Oxford, 1989), p. 185.
\bibitem{Laplace} H. Stehfest, Comm. ACM {\bf 13}, 47 (1970).
\bibitem{SL} For a similar approach see also
G. Szamel and J. A. Leegwater, Phys. Rev. A {\bf 46}, 5012 (1992).
\bibitem{SCremark} A similar additional variable was used in the 
more advanced version of the CS theory \cite{CS2}. They, however, used disorder
averaged quantities throughout.
\bibitem{comment5} It can be argued that the neglected correlations
do not contribute in the thermodynamic limit.
\bibitem{Kawasaki2} K. Kawasaki, Physica A {\bf 215}, 61 (1995).
\end{references}
\end{document}